\documentstyle[aps,prl,multicol,epsf]{revtex}
\begin{document}
\draft 
\title{Feynman's Propagator Applied to Network Models of Localization} 
\author{H. Mathur} 
\address{Department of Physics,
Case Western Reserve University, Cleveland, Ohio
44106-7079} 

\date{\today} 
\maketitle


\begin{abstract}
Network models of dirty electronic systems are mapped onto an 
interacting field theory of lower dimensionality by interpreting
one space dimension as time. This is accomplished via Feynman's
interpretation of anti-particles as particles moving backwards 
in time. The method developed maps calculation of the moments of
the Landauer conductance onto calculation of correlation functions
of an interacting field theory of bosons and fermions. The resulting
field theories are supersymmetric and closely related to the 
supersymmetric
spin-chain representations of network models recently discussed by 
various authors. As an application of the method, the two-edge 
Chalker-Coddington model is shown to be Anderson localized, and a 
delocalization
transition in a related two-edge network model (recently discussed by 
Balents and Fisher) is studied by calculation of the average Landauer
conductance. 
\end{abstract} 
\pacs{PACS: }

\begin{multicols}{2} 

\input epsf

\section{Introduction}

Dirty electronic systems exhibit a variety of interesting
phases and transitions in their transport properties at low
temperature. For example consider two dimensional electrons 
moving in a random potential and a strong
perpendicular field, as in a quantum Hall experiment. Generically
the wavefunctions are localized, with tails that decay 
over a length scale called the localization length,
which leads to insulating behaviour at low temperature. 
However if the parameters
(e.g., the magnetic field or electron density) are tuned 
to special isolated values the localization length diverges.
Such delocalization transitions were postulated shortly after
the discovery of the quantum Hall effect and invoked in 
order to explain it \cite{girvin}. Since then impressive strides 
have been
made in experimental characterization of these transitions 
\cite{sondhi};
but a satisfactory theoretical understanding has not yet
emerged \cite{huckestein}. 

An important theoretical advance was made by Chalker and Coddington
who introduced a simplified network model of the quantum Hall
transition \cite{chalker}. Numerical studies show this model 
produces the 
same universal behaviour at the delocalization transition
as more literal (and more complicated) models of the quantum 
Hall system \cite{huckestein}. Because the network model is 
relatively simple 
and is based on a clear physical picture of the transition, it 
seems a promising starting point for a controlled approximate
analysis of the transition.

More recently it has been observed that the random bond
Ising model is closely related to a variation on the network
model \cite{cho}. Progress in analysis of network models is
therefore desirable from this point of view also.

Conductance is a sensitive probe of delocalization. The purpose
of this paper is to introduce a technique suitable for calculation
of the conductance of network models. Following Landauer we imagine
electrons are injected from the source into the sample where they
undergo multiple scattering \cite{aaron}. Eventually
an electron may either get scattered forward into 
the drain or it may get scattered backward into the source (see
fig 1). We wish to calculate the probability of forward scattering.

It is often fruitful in statistical mechanics to 
map a problem onto a quantum field theory of lower dimensionality
by reinterpreting one space dimension as time. In applying 
that strategy here it becomes neccessary to take into account the
fact that electrons will then appear to move both forwards
and backwards in ``time''. 


\epsfxsize=3.5in \epsfbox{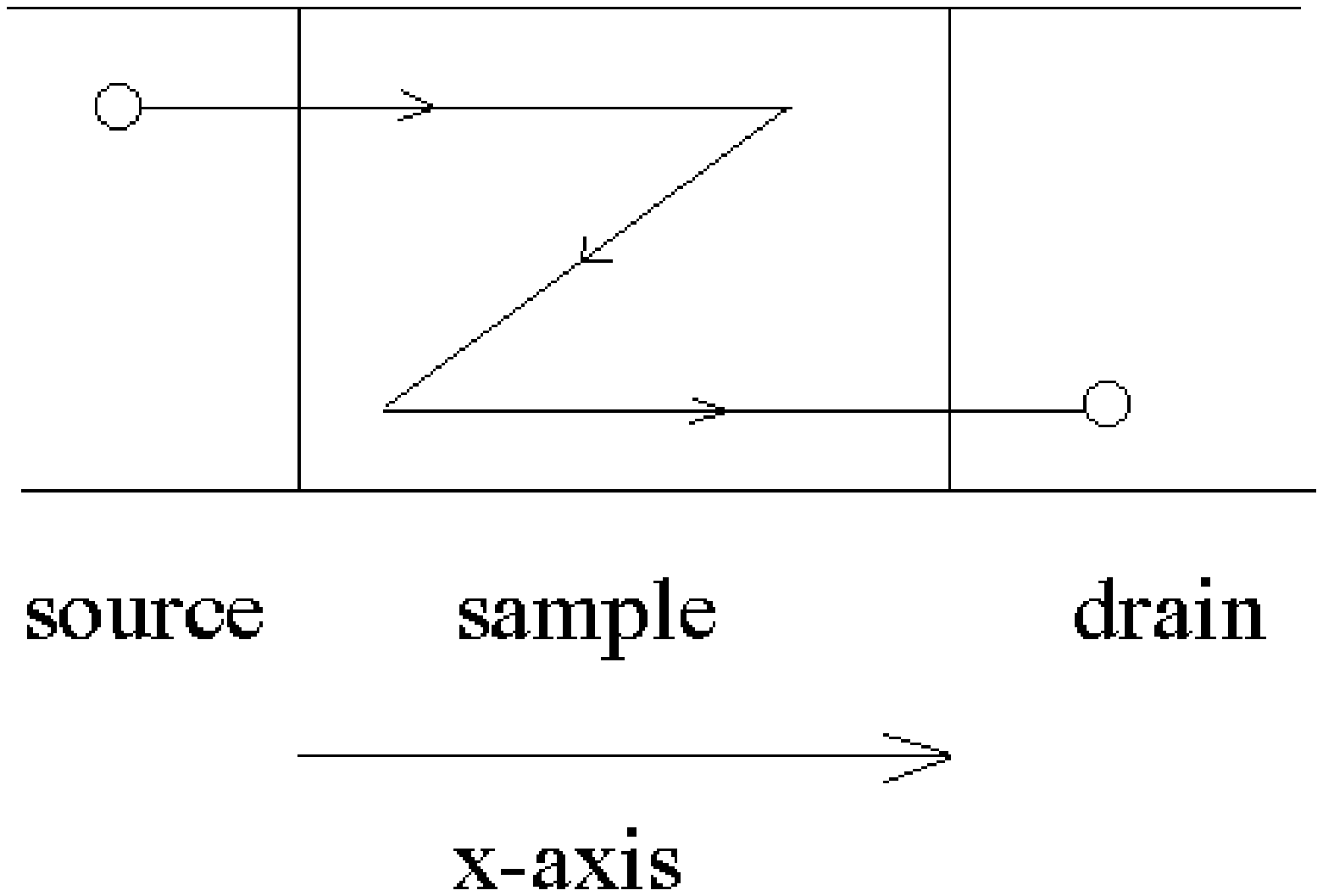}

\figure{Figure 1. Landauer's picture: The electron undergoes multiple
scattering in the sample and is eventually scattered forward into the
drain. In the field theory representation this corresponds to a 
process
involving pair creation and annhilation.}

A similar situation is encountered in quantum
electrodynamics: in a famous paper, {\em Theory of Positrons},
Feynman showed that it is possible to regard positrons as 
electrons moving backwards in time \cite{feynman}. Following 
Feynman it is
possible here to interpret the x-axis in fig 1 as time provided we
regard an electron moving to the left as a sort
of anti-particle of an electron moving to the right. For example
the process in which a right moving electron is scattered 
to the left could be regarded as a process in which the particle
meets it's anti-particle leading to their mutual annhilation.
From this perspective, the simple process depicted in fig 1
where an electron zig-zags through the sample 
could be regarded as a process in which a particle-antiparticle
pair is created (at the earlier kink in the trajectory) and
annhilated (at the later kink). 

Based on this interpretation, it is possible to
map a two-dimensional network model onto a one-dimensional
field theory of particles
and anti-particles. The statistics of the particles may be taken
to be either Bose or Fermi. For technical reasons it is most
convenient to introduce both species. Thus calculation of the
conductance is mapped onto calculation of the correlation functions
of an interacting field theory of bosons and fermions. 

For clarity we illustrate the method on two-edge
network models which are essentially one-dimensional. They 
map onto quantum mechanics
problems of interacting bosons and fermions---a zero-dimensional
field theory. Generalization of the mapping
to a two-dimensional network model
is straightforward. Indeed the method should be much more broadly
applicable to dirty electronic systems. Since an important
obstacle to non-perturbative analysis of random systems has been the
lack of suitable representations of the problem, it is hoped
this method may prove useful. The present method is closely
related to supersymmetric spin-chain representations of network
models that have recently been discussed by several authors
\cite{zirnbauer,leon1,ilya1,ilya2,kondev,leon2}. 

As an application of the method we analyse two 
different network models. 
The first is essentially a one-dimensional Chalker-Coddington
model which exhibits Anderson localization: for a large enough
sample the zero temperature conductance is found to decay 
exponentially
with sample size. 

The second model was recently discussed in context of quantum 
transport
by Balents and M.P.A. Fisher \cite{leon2}. 
It is related to a model of glass first
studied by Dyson \cite{dyson} and to a special version of the 
random bond Ising
model introduced by McCoy and Wu \cite{mccoy}. 
An enlightening discussion of these
connections, with references, is given in the paper of Balents 
and M.P.A.
Fisher. This model is known to have a critical point at a special
value of its parameters which is of considerable interest as a simple
example of a random quantum critical point. Much is known about
this model and its relations, particularly through the application
of real space renormalization group methods by D.S. Fisher 
\cite{daniel}. 
In their
recent paper Balents and M.P.A. Fisher have applied supersymmetry
methods to calculate the exact two-parameter scaling function of the
Green's function of this model. Here we shall study it's 
delocalization
transition by calculating the conductance. A summary of our results
for this model is given in section IVC.

\section{Field Theory Formulation}

In this section a simple two-edge network model is introduced
and it's Landauer conductance defined. The main result is an exact
formal expression for the Landauer conductance, eq (42-44), within
a field theory formulation of the problem.  

\subsection{Landauer Conductance}

The models considered in this paper consist of two counter-propagating
``edge states'' coupled by tunnelling. Along each one-dimensional 
edge electrons
can propagate in only one direction (see fig 2). The electron wave
function has two components: $ \psi_{+} (x)$, the amplitude to be on
the right moving edge at position $x$, and $ \psi_{-}(x) $, the 
amplitude
to be on the left moving edge at position $x$. The time independent
Schr\"{o}dinger equation governing this model is 
\begin{equation}
\left(  \begin{array}{cc}
- i \frac{\partial}{\partial x} & m(x) \\
m^{*}(x) & i \frac{\partial}{\partial x}
\end{array}
\right)
\left( \begin{array}{c}
\psi_{+} \\
\psi_{-}
\end{array} \right)
= E \left( \begin{array}{c}
\psi_{+} \\
\psi_{-} 
\end{array} \right).
\end{equation}
The tunnelling amplitude $m(x)$ is some given function. Eventually
we will take tunnelling to be a random process and will be interested
in performing averages over an ensemble of different 
realizations of $m(x)$ 
with statistics to be described below. We are interested in solutions 
at some fixed energy $E$ (the fermi energy). 

For later reference, it is useful to rewrite eq (1) in the form
\begin{equation}
- i \frac{ \partial }{ \partial x } 
\left( \begin{array}{c}
\psi_{+} \\
\psi_{-} 
\end{array}
\right)
= \left( \begin{array}{cc}
E & - m(x) \\
m^{*}(x) & -E 
\end{array}
\right)
\left( \begin{array}{c}
\psi_{+} \\
\psi_{-}
\end{array} \right)
\end{equation}



\epsfxsize=3.5in \epsfbox{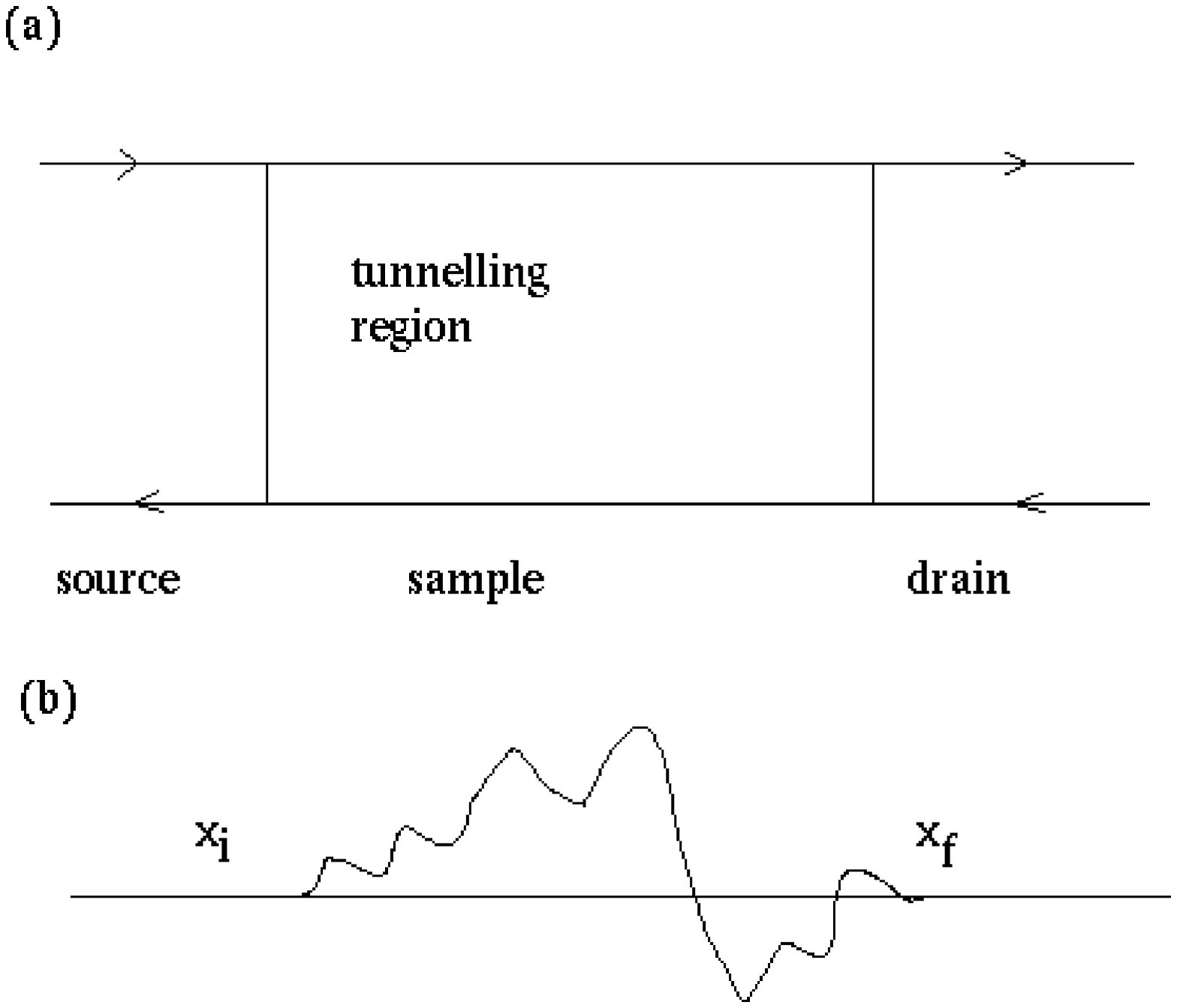}

\figure{Figure 2 (a) A two edge network model. (b) Schematic
picture of the random function $ m(x) $. }

An obvious generalization of this model is to consider 2$N$
counter-propagating edge states. With suitable statistical 
assumptions 
about the tunnelling these generalizations lead to the 
Chalker-Coddington
model of the quantum Hall transition \cite{chalker} 
or the random bond Ising model \cite{cho}.
The model studied here can be considered a special anisotropic case
in which tunnelling between alternate pairs of edges has been turned
off. 

Following Landauer's method \cite{aaron}, the model is separated 
into the sample
(region between $ x_{i} $ and $x_{f}$) and the probes (regions to
the left of $x_{i}$ and to the right of $x_{f}$); see fig 2. The
left probe is called the source; the right probe, the drain. In 
contrast to the sample the probes are assumed to be disorder free.
$ m(x) $ is therefore random only for $ x_{i} < x < x_{f} $. It
is constant in the probes. For simplicity we shall assume that
the edges are uncoupled in the probes 
(that is, $ m(x) = 0$)\footnote{
It is not essential for our approach to assume $ m(x) = 0 $ in the 
probes.
However, it simplifies the analysis. It would be expected that the 
transport
properties of the sample should not be modified by the nature of the 
probes.
This can be checked explicitly under some circumstances.}.
This is shown schematically in fig 2 (b).  

The wave eq (2) is now trivially soluble in the probes. We find
\begin{eqnarray}
\left( \begin{array}{c}
\psi_{+} \\
\psi_{-} 
\end{array}
\right) & = & 
\begin{array}{cc}
\left( \begin{array}{c}
\alpha \exp i E x \\
\beta \exp - i E x
\end{array}
\right) & {\rm source} 
\end{array} 
\nonumber \\
 & = & \begin{array}{cc}
\left( \begin{array}{c}
\gamma \exp i E x \\
\delta \exp - i E x 
\end{array}
\right) & {\rm drain}
\end{array}. 
\end{eqnarray}
For a given realization of $m(x)$ the connection between 
$ (\gamma, \delta) $ and
$(\alpha, \beta)$ can be found by integrating eq (2) across the sample
\begin{equation}
\left( \begin{array}{c}
\gamma \\
\delta
\end{array}
\right) 
= T \left( \begin{array}{c}
\alpha \\
\beta
\end{array}
\right).
\end{equation}
$ T $ is a 2$\times$2 matrix called the transfer matrix of the sample.
The transfer matrix is the focus of numerical studies of network 
models
\cite{huckestein}.

Here we are interested in solving eq (2) subject to scattering 
boundary
conditions in the probes so that
\begin{equation}
\alpha \rightarrow 1, \hspace{2mm} \beta \rightarrow r; 
\hspace{3mm} \gamma \rightarrow t, \hspace{2mm} \delta \rightarrow
0.
\end{equation}
According to Landauer the zero-temperature
conductance of the sample (at Fermi energy $E$) is
given by
\begin{equation}
g = \frac{ e^2 }{ h } | t |^2.
\end{equation} 

Returning to eq (2) it is useful to regard $ x \rightarrow 
{\rm time}$ and to
interpret it as the {\em time-dependent} Schr\"{o}dinger equation of
a two-level system. An awkward feature of such a reinterpretation is 
that
the 2$\times$2 ``Hamiltonian'' matrix in eq (2) is not 
Hermitean\footnote{
Perhaps this is a good place to emphasize that in this paper 
although
the ``Hamiltonian'' is frequently non-Hermitean, it is always a 
well defined
operator in a well defined Hilbert space. There are no difficulties of 
interpretation as our ``Hamiltonian'' is merely used as a calculational
device. The true Schr\"{o}dinger equation of the model, eq (1), is governed
by a Hermitean Hamiltonian.}. Sadly this feature will persist through much
of our analysis and appears to be more generally pervasive 
\cite{zirnbauer,kondev,leon2}.

We can now imagine calculating the retarded propagator for eq (2). Given
the wave function (values of $ \psi_{+} $ and $ \psi_{-} $) at a particular
$x-$slice, the retarded propagator gives the wave function for ``future''
values of $x$. For example, if the wave function is known at $ x_{i} $,
the retarded propagator gives the wave function inside the sample (and
beyond, in the drain). This is in very much the same spirit as the
transfer matrix (indeed the retarded propagator {\em is} the transfer
matrix for a special value of it's arguments). It is not helpful for
calculating the scattering amplitudes.

While solving eq (2) subject to scattering boundary conditions we do
not know the complete wave function for any $ x-$slice. Instead we
know the positive frequency component $ \psi_{+} = 1$ at the earlier
slice $x_{i}$ and we know the negative frequency component $\psi_{-} = 0$
at a later slice $x_{f}$. We would like to reconstruct the wave function
between the slices using this information; or at least we would like
to know $ \psi_{+} ( x \rightarrow x_{f} ) = t $. 

In {\em Theory of Positrons} Feynman showed precisely
this would be achieved by a modification of the retarded propagator, now
known as the Feynman propagator \cite{feynman}. 
Feynman derived an integral equation
obeyed by his propagator and a perturbation series for it; this was
the main focus of his paper. Peripherally, in two Appendixes, he showed
the same results could be deduced from a second quantized theory of
particles and anti-particles. This development will now be used to provide
a field theory representation of the two edge network model (eq 2) \footnote{
The next two subsections follow Feynman very closely. For this reason
we have given them the same titles as the corresponding appendixes in
Feynman's paper}. 

\subsection{Deduction From Second Quantization} 

The purpose of this section is to derive a second quantized Hamiltonian
from which one can calculate the Feynman propagator and scattering amplitude
$t$ of eq (2).

\subsubsection{Fermion Representation}

Return to the interpretation of eq (2) as a time dependent two level system.
Introduce $ a_{+}^{\dagger} $ which creates the right moving state 
$ \left( \begin{array}{c} 1 \\ 0 \end{array}
\right) $; $ a_{-}^{\dagger} $ which creates the left moving state 
$\left( \begin{array}{c} 0 \\ 1 \end{array}
\right) $; and $ a_{+} $ and $ a_{-} $, the corresponding destruction operators. The
$ a$'s obey fermi (anti-)commutation rules. Also introduce the ``time''-dependent
Hamiltonian
\begin{equation}
H (x) = ( a_{+}^{\dagger} a_{-}^{\dagger} )
\left( \begin{array}{cc}
E & -m(x) \\
m^{*}(x) & -E 
\end{array}
\right)
\left( \begin{array}{c} 
a_{+} \\
a_{-} 
\end{array}
\right).
\end{equation}
This Hamiltonian is not useful for our purpose. It can be shown to generate
the retarded propagator.

Instead we must now introduce $c^R$ fermions, related to the $ a $ fermions
via the particle-hole transformation
\begin{equation}
\begin{array}{cc}
c_{+}^{R \dagger} = a_{+}^{\dagger}; & c_{-}^{R \dagger} = a_{-} ; \\
c^{R}_{+} = a_{+} ; & c^{R}_{-} = a_{-}^{\dagger}.
\end{array}
\end{equation}
The superscript $R$ is superfluous for the moment; its function
will become apparent later. 
In terms of the $c^R$ fermions the Hamiltonian is
\begin{equation}
H_{F}^{R} (x) = E( c_{+}^{R \dagger} c_{+}^{R} + c_{-}^{R \dagger}
c_{-}^{R} ) - m(x) c_{+}^{R \dagger} c_{-}^{R \dagger}
+ m^{*}(x) c_{-}^{R} c_{+}^{R}. 
\end{equation}
The scattering amplitude $t$ can be computed straightforwardly using
this Hamiltonian; see eq (19) below.
In the remainder of this subsubsection that result will be 
derived.

Note that $H_{F}$ is non-Hermitean---this is traceable to the non-Hermiticity
of the ``Hamiltonian'' of eq (2). The reader troubled by this non-Hermiticity
should again read footnote 2. Note also that $H_{F}$ does not conserve
the total number of $c-$fermions. Instead we may regard the $c_{-}$fermion as
the ``anti-particle'' of the $c_{+}$. What is conserved then is the total
``charge''---the difference in the number of particles and anti-particles,
given by
\begin{equation}
Q_{F} = c_{+}^{R \dagger} c^{R}_{+} - c_{-}^{R \dagger} c^{R}_{-}.
\end{equation}

The S-matrix corresponding to $H_{F}^{R}$ is the solution to
\begin{equation}
- i \frac{\partial}{\partial x} S^{R}_{F} (x, x_{i}) = 
H_{F}^{R} (x) S^{R}_{F}(x, x_{i})
\end{equation}
subject to the initial condition $S_{F}^{R}(x_{i}, x_{i}) = 1$.

Turning to the derivation of eq (19), we first obtain a useful
formula, following Feynman.
Let $( e_{+}(x), e_{-}(x) )$ be a solution to the wave eq (2). 
Construct the operator $ \hat{F} \equiv e_{+}( x_{i} ) c_{+}^{R 
\dagger} + e_{-} (x_{i}) c_{-}^{R}$. $\hat{F}$ is transformed by the 
$S-$matrix as follows:
\begin{equation}
S^{R}_{F} (x, x_{i}) \hat{F} S_{F}^{R -1} (x, x_{i}) =
e_{+} (x) c^{R \dagger}_{+} + e_{-}(x) c^{R}_{-}.
\end{equation}  
This is Feynman's formula. To prove it, regard eq (12) as
an ansatz; in other words assume that $S \hat{F} S^{-1}$
is of this form with $ e_{+}(x) $ and $ e_{-}(x)$ some
suitable functions. Upon differentiation with respect to
$x$, it will be seen that the ansatz is consistent provided
$ ( e_{+}, e_{-} ) $ obey the wave eq (2). 

Carrying out this plan, we note
\begin{equation}
- i \frac{\partial}{\partial x} S_{F}^{R -1} (x, x_{i})
= - S_{F}^{R -1} (x, x_{i}) H_{F}^{R} (x)
\end{equation}
obtained by differentiating $ S_{F}^{R -1} S_{F}^{R} = 1$
and using the evolution eq (11) for the $S-$matrix. 
Eq (11), (12) and (13) together yield
\begin{eqnarray}
- i \frac{\partial}{\partial x} \left( S_{F}^{R} \hat{F} S_{F}^{R -1}
\right) & = & [ H_{F}, S_{F}^{R} \hat{F} S_{F}^{R -1} ] \nonumber \\
 & = & e_{+}(x) [ H_{F}, c_{+}^{R \dagger} ] + e_{-} (x) [ H_{F}, c_{-}^{R} ].
 \nonumber \\
 & &
\end{eqnarray}
The relevant commutators are
\begin{eqnarray}
[ H_{F}, c_{+}^{R \dagger} ] & = & E c_{+}^{R \dagger} + m^{*} c_{-}^{R} 
\nonumber \\
{[ H_{F}, c_{-}^{R} ]} & = & - E c_{-}^{R} - m c_{+}^{R \dagger}
\end{eqnarray}
Hence the derivative of the left side of the ansatz eq (12) is 
\begin{eqnarray}
- i \frac{\partial}{\partial x} \left( S_{F}^{R} \hat{F} 
S_{F}^{R -1} \right) & =  & c_{+}^{R \dagger} [ E e_{+} (x) - m(x) e_{-} (x) ]
 \nonumber \\
 & & + c_{-}^{R} [ m^{*} (x) e_{+}(x) - E e_{-} (x) ]. 
\end{eqnarray}  
On comparing with the corresponding $x-$derivative of the right hand side
of the ansatz it is seen that indeed $ (e_{+}, e_{-}) $ must obey the wave
eq (2); this completes the proof of Feynman's formula eq (12).

Next let $(e_{+}, e_{-})$ be the special solution that obeys scattering
boundary conditions:
\begin{equation}
\begin{array}{cc}
e_{+}(x_{i}) = 1; & e_{-}(x_{i}) = r; \\
e_{+}(x_{f}) = t; & e_{-}(x_{f}) = 0. 
\end{array}
\end{equation}
This leads to the {\em scattering} formula
\begin{equation}
\begin{array}{c}
S_{F}^{R} (x_{f}, x_{i}) ( c_{+}^{R \dagger} + r c_{-}^{R} )
S_{F}^{R -1} (x_{f}, x_{i})  = t c_{+}^{R \dagger} \\ [5mm]
\Rightarrow S_{F}^{R} (x_{f}. x_{i}) c_{+}^{R \dagger} = 
t c_{+}^{R \dagger} S_{F}^{R} (x_{f}, x_{i}) - r S_{F}^{R} (x_{f}, x_{i})
c_{-}^{R},
\end{array}
\end{equation}
a useful special case of Feynman's formula, eq (12).

Finally analyse the matrix element  
$<0| c_{+}^{R} S_{F}^{R} (x_{f}, x_{i}) c_{+}^{R \dagger} |0>$;
here $|0>$ is the vacuum for the $c-$fermions. Use of the scattering
formula, eq (18), reveals
\begin{equation}
t = \frac{ <0| c_{+}^{R} S_{F}^{R} (x_{f}, x_{i}) c_{+}^{R \dagger} |0>}
{<0|S_{F}^{R} (x_{f}, x_{i}) |0>}.
\end{equation}
Eq (19) shows that the scattering amplitude $t$ can be computed by
studying the evolution of the single particle state $c_{+}^{R \dagger}
|0>$ under the Hamiltonian of eq (9). Note that the vacuum amplitude
in the denominator of eq (19) is not trivial. In Feynman's words, ``It
differs from unity because, for example, a pair could be created which 
eventually annhilates itself''.

Eq (19) is the main result of this subsubsection. By itself, eq (19)
is not especially helpful. Our eventual purpose is to compute averages
over the random tunnelling process $m(x)$. Since $m(x)$ appears in
both numerator and denominator of eq (19), this form is not particularly
well adapted for averaging.

\subsubsection{Boson Representation}

Tracing through the derivation of eq (19) it is seen that the Fermi
statistics of the $c^{R}$ particles plays no crucial role; only the
commutators of eq (15) are essential. Hence we may replace the
$c^{R}$ fermions with $b^{R}$ bosons governed by the Hamiltonian
\begin{equation}
H_{B}^{R} (x) = E( b_{+}^{R \dagger} b_{+}^{R} + b_{-}^{R \dagger}
b_{-}^{R} ) +  m(x) b_{+}^{R \dagger} b_{-}^{R \dagger} 
+ m^{*}(x) b_{-}^{R} b_{+}^{R}.
\end{equation}
Apart from the replacement $ c^{R} \rightarrow b^{R} $, this differs
from eq (9) in the sign of the pair creation term (the term proportional
to $b_{+}^{R \dagger} b_{-}^{R \dagger}$). This sign change ensures
commutation relations of the desired form similar to eq (15):
\begin{eqnarray}
[ H_{B}^{R}, b_{+}^{R \dagger} ] & = & E b_{+}^{R \dagger} + 
m^{*}(x) b_{-}^{R},
\nonumber \\
{[ H_{B}^{R}, b_{-}^{R \dagger} ]} & = & - E b_{-}^{R} - m(x) b_{+}^{R \dagger}.
\end{eqnarray}

Note that due to the sign change in the pair creation term the boson
Hamiltonian, eq (20), is actually Hermitean. Again the $b_{-}$ boson
may be regarded as the anti-particle of the $b_{+}$. The Hamiltonian
conserves the difference in the number of particles and anti-particles,
that is, the charge
\begin{equation}
Q_{B} = b_{+}^{R \dagger} b_{+}^{R} - b_{-}^{R \dagger} b_{-}^{R}.
\end{equation}

The entire analysis of the preceding subsubsection can now be carried
over essentially unchanged. The S-matrix corresponding to $H_{B}^{R}$ is 
the solution to
\begin{equation}
- i \frac{\partial}{\partial x} S^{R}_{B} (x, x_{i}) = 
H_{B}^{R} (x) S^{R}_{B}(x, x_{i})
\end{equation}
subject to the initial condition $ S_{B}^{R}( x_{i}, x_{i} ) = 1 $. 
Feynman's formula, eq (12), and the scattering formula, eq (18),
apply upon making the replacement $ c^{R} \rightarrow b^{R} $. 
Using the scattering formula we can deduce
\begin{equation}
t = \frac{ <0| b_{+}^{R} S_{B}^{R} (x_{f}, x_{i}) b_{+}^{R \dagger} |0>}
{<0|S_{B}^{R} (x_{f}, x_{i}) |0>}
\end{equation}
---the bosonic analogue of eq (19).

\subsection{Analysis of the Vacuum Problem}

Remarkably, the vacuum amplitudes for the bosons and fermions cancel;
that is,
\begin{equation}
<0|S_{F}^{R}(x_{f},x_{i})|0> <0|S_{B}^{R}(x_{f},x_{i})|0> = 1.
\end{equation}
We have previously derived a fermionic and bosonic expression for
the scattering amplitude, neither well suited for performing an
average over the random tunnelling process, $m(x)$. Eq (25) will
enable us to weld these expressions into a form suitable for averaging.
We return to this point in section E below. Here we focus on proving 
eq (25).

To calculate the vacuum amplitude, following Feynman, let us analyse
a series of problems that interpolate smoothly between a soluble limit
and the problem we want to solve. Introduce a truncated problem for which
$ m(x)$ is left unchanged for $ x_{0} < x < x_{f} $ and is set equal to
zero for $ x_{i} < x < x_{0} $. By varying $ x_{0} $ we obtain the
desired series of problems. 
Evidently $ x_{0} = x_{i} $ is the case we wish to solve. On the other
hand, if $ x_{0} = x_{f} $, there is no tunnelling and the problem is 
trivially soluble.

Denote the scattering coefficients for the truncated problem $ r (x_{0})$
and $ t (x_{0}) $. For the soluble case, $ r(x_{f}) = 0 $ and $ t(x_{f}) = 1 $.

The fermion S-matrix for the truncated problem will be written as
$ S_{x_{0}}^{RF} (x, x_{0}) $. It obeys 
\begin{equation}
- i \frac{ \partial }{ \partial x} S_{x_{0}}^{RF} (x, x_{0}) = H_{F}^{R} (x)
S_{x_{0}}^{RF} (x, x_{0})
\end{equation}
with $ S_{x_{0}}^{RF} (x_{0}, x_{0}) = 1 $. The truncated boson S-matrix is
similarly defined.

The vacuum amplitude for the truncated fermion problem will be denoted
$ C_{F}^{R} (x_{0}) $. Thus
\begin{equation}
C_{F}^{R} (x_{0}) = <0| S_{x_{0}}^{RF} (x_{f}, x_{0}) |0>
\end{equation}
Analogous expressions can be written for the bosonic case.
When $ x_{0} = x_{f} $ the vacuum amplitude is unity for both
bosons and fermions.

The strategy now is to study the variation of $ C(x_{0}) $ with $ x_{0} $
by writing a differential equation for it. A formal solution of 
the differential
equation can be obtained which will suffice to prove the cancellation of the 
vacuum amplitude.

For this purpose it is useful to write the solution to eq (26) as a 
formal series
\begin{eqnarray}
S_{x_{0}}^{RF} (x_{f}, x_{0}) & = & 1 + i \int_{x_{0}}^{x_{f}} d x_{1}
H_{F}^{R} (x_{1}) \nonumber \\
 & & + i^2 \int_{x_{0}}^{x_{f}} d x_{1} \int_{x_{0}}^{x_{1}} d x_{2}
 H_{F}^{R}(x_{1}) H_{F}^{R} (x_{2}) \nonumber \\
 & & + \ldots
 \end{eqnarray}
By differentiation of this series it follows
\begin{equation}
- i \frac{ \partial }{ \partial x_{0} } S_{x_{0}}^{RF} (x_{f}, x_{0}) = 
-  S_{x_{0}}^{RF} (x_{f}, x_{0}) H_{F}^{R}(x_{0}).
\end{equation}
Taking the vacuum expectation of eq (29) obtain
\begin{eqnarray}
- i  \frac{ \partial }{ \partial x_{0} } C_{x_{0}}^{RF} & = & 
- <0| S_{x_{0}}^{RF} (x_{f}, x_{0}) H_{F}^{R} (x_{0}) |0>; \nonumber \\
H_{F}^{R}(x_{0}) & = & 
- m(x_{0}) c_{+}^{R \dagger} c_{-}^{R \dagger} + {\rm others } 
\end{eqnarray}
In eq (30) we have explicitly written only that term of $H_{F}^{R}$ (eq 9) which
does not annhilate the vacuum.

According to the scattering formula, eq (18),
\begin{eqnarray}
S_{x_{0}}^{RF} (x_{f}, x_{0}) c_{+}^{R \dagger} & = & t(x_{0}) c_{+}^{R \dagger}
S_{x_{0}}^{RF} (x_{f}, x_{0}) \nonumber \\
& & - r( x_{0} ) S_{x_{0}}^{RF} (x_{f}, x_{0})
c_{-}^{R}.
\end{eqnarray}
Applying this to the matrix element in eq (30) yields the differential
equation obeyed by $ C_{F}^{R} (x_{0})$:
\begin{equation}
- i \frac{ \partial }{\partial x_{0} } C_{F}^{R} (x_{0}) = - m (x_{0}) r (x_{0})
C_{F}^{R} (x_{0}).
\end{equation}
We seek a solution subject to $ C_{F}^{R} (x_{f}) = 1$.

Define $ L( x_{0} )$ as the solution of
\begin{equation}
- i \frac{ \partial }{ \partial x_{0} } L( x_{0} ) = - m(x_{0}) r (x_{0}) 
\end{equation}
with the initial condition $ L( x_{f} ) = 0 $. Given $ m(x) $ one can
imagine solving the truncated problem for various values of $ x_{0} $
and, in principle, computing $ L(x_{0})$. The solution of eq (32) is then
\begin{equation}
C_{F}^{R} (x_{0}) = \exp L(x_{0}).
\end{equation}
Note that this solution obeys the differential equation (32) {\em and} the
initial condition $ C_{F}^{R}(x_{f}) = 1$. 

The same analysis leads to the boson vacuum amplitude
\begin{equation}
C_{B}^{R} (x_{0}) = \exp - L(x_{0}).
\end{equation}
The sign change is traceable to the sign difference in the pair creation
terms of the Fermi and Bose Hamiltonians.
Hence $C_{F}^{R} (x_{0}) C_{B}^{R} (x_{0}) = 1 $ for all $ x_{0} $ and
in particular for $ x_{0} = x_{i} $, which establishes the desired result,
eq (25).

It is interesting that Feynman appears to have devoted some effort
to interpreting Dirac particles as bosons even though it violated
the spin-statistics theorem. In particular, the result of eq (25),
which is of no obvious value in quantum electrodynamics, is derived in
the {\em Theory of Positrons}. 

\subsection{Conjugate Amplitude}

The Landauer conductance is given by $ |t|^2 $. Thus expressions for
$ t^* $ analogous to eq (19) and (24) will be needed. To obtain them
consider the equation conjugate to eq (2):
\begin{equation}
- i \frac{ \partial }{ \partial x } 
\left( \begin{array}{c}
\phi_{+} \\
\phi_{-} 
\end{array}
\right)
= \left( \begin{array}{cc}
- E & m^{*}(x) \\
- m(x) & E 
\end{array}
\right)
\left( \begin{array}{c}
\phi_{+} \\
\phi_{-}
\end{array} \right)
\end{equation}
Eq (36) is constructed to have the property that if $(e_{+}, e_{-})$ is
a solution to eq (2), then the complex conjugate $(e_{+}^{*}, e_{-}^{*})$
is a solution to eq (36). Hence if eq (36) is solved subject to scattering
boundary conditions, the scattering amplitude will be $ t^{*} $.

On the other hand, direct comparison reveals that eq (2) is transformed
into eq (36) by the replacements $ E \rightarrow - E, m \rightarrow - m^*,
m^* \rightarrow - m$. Hence introduce $ c^A$ fermions governed by the
Hamiltonian
\begin{equation}
H_{F}^{A}(x) = - E ( c_{+}^{A \dagger} c_{+}^{A} + c_{-}^{A \dagger}
c_{-}^{A} )
+ m^{*}(x) c_{+}^{A \dagger} c_{-}^{A \dagger}
- m(x) c_{-}^{A} c_{+}^{A}.
\end{equation}
This Hamiltonian is obtained from eq (9) by making the replacements 
indicated above. By the reasoning that lead from eq (9) to (19),
$ H_{F}^{A} $ generates the conjugate amplitude via
\begin{equation}
t^{*} = \frac{ <0| c_{+}^{A} S_{F}^{A} (x_{f}, x_{i}) c_{+}^{A \dagger} |0>}
{<0|S_{F}^{A} (x_{f}, x_{i}) |0>}
\end{equation}
where $ S_{F}^{A} (x_{f}, x_{i}) $ is the S-matrix corresponding to 
$ H_{F}^{A} $.

Similarly, one can write
\begin{equation}
t^{*} = \frac{ <0| b_{+}^{A} S_{B}^{A} (x_{f}, x_{i}) b_{+}^{A \dagger} |0>}
{<0|S_{B}^{A} (x_{f}, x_{i}) |0>}
\end{equation} 
where $ S_{B}^{A} (x_{f}, x_{i}) $ is the S-matrix of the boson Hamiltonian
\begin{equation}
H_{B}^{A} = - E ( b_{+}^{A \dagger} b_{+}^{A} + b_{-}^{A \dagger}
b_{-}^{A} ) - m^{*}(x) b_{+}^{A \dagger} b_{-}^{A \dagger}
- m(x) b_{-}^{A} b_{+}^{A}.
\end{equation}

Finally note that the vacuum amplitudes for the A bosons and fermions
cancel, as they do for their R counterparts:
\begin{equation}
<0|S_{F}^{A} (x_{f}, x_{i}) |0> <0|S_{B}^{A} (x_{f}, x_{i}) |0> = 1.
\end{equation}

\subsection{Supersymmetry}

The results of the previous subsections can now be assembled into an
expression for the Landauer conductance suitable for averaging over
the random tunnelling process $ m(x) $. Simultaneously introduce $ c^R$
and $ c^A $ fermions and $ b^R $ and $ b^A $ bosons governed by the
total Hamiltonian
\begin{eqnarray}
H_{SUSY} (x) & = & H_{F}^{R} + H_{F}^{A} + H_{B}^{R} + H_{B}^{A}
\nonumber \\
 & = & E \hat{M} + m(x) \hat{A} + m^{*}(x) \hat{B}
\end{eqnarray}
Here $ \hat{M} \equiv ( c_{+}^{R \dagger} c_{+}^{R} + c_{-}^{R \dagger}
c_{-}^{R} - R \rightarrow A ) + ( b_{+}^{R \dagger} b_{+}^{R} +
b_{-}^{R \dagger} b_{-}^{R} - R \rightarrow A )$ and
$ \hat{A} \equiv  - c_{+}^{R \dagger} c_{-}^{R \dagger} 
- c_{-}^{A} c_{+}^{A} + b_{+}^{R \dagger} b_{-}^{R \dagger}
- b_{-}^{A} b_{+}^{A}  $ and
$\hat{B} \equiv c_{-}^{R} c_{+}^{R} + c_{+}^{A \dagger} c_{-}^{A \dagger}
+ b_{-}^{R} b_{+}^{R} - b_{+}^{A \dagger} b_{-}^{A \dagger} $.

As usual the corresponding S-matrix obeys
\begin{equation}
- i \frac{ \partial }{ \partial x } S_{SUSY} (x, x_{i}) = 
H_{SUSY} (x) S_{SUSY} (x, x_{i})
\end{equation}
and the initial condition
$ S_{SUSY} (x_{i}, x_{i}) = 1$.

The Landauer conductance can be calculated using
\begin{equation}
|t|^2 = <0| c_{+}^{A} c_{+}^{R} S_{SUSY}( x_{f}, x_{i} ) c_{+}^{R \dagger}
c_{+}^{A \dagger} |0>.
\end{equation}
To see this note that since the different particle species don't interact,
$S_{SUSY} = S_{F}^{R} S_{F}^{A} S_{B}^{R} S_{B}^{A} $; hence the matrix
element in eq (44) decouples into 
\begin{displaymath}
<0| C_{+}^{R} S_{F}^{R} c_{+}^{R \dagger} 
|0> <0| c_{+}^{A} S_{F}^{A} c_{+}^{A \dagger} |0> <0| S_{B}^{R} |0>
<0| S_{B}^{A} |0>.
\end{displaymath} 
Using eq (19), (38), (25) and (41), this product
is easily seen to be $ |t|^2$. 

Eq (44) is our principal tool to analyse the network model.
It shows that the conductance can be calculated by following
the evolution of a two fermion state under the Hamiltonian
$H_{SUSY}$. The principal advantage of this expression is due to
the absence of a denominator which makes it well suited for 
averaging over the random tunnelling process $m(x)$. The averaging
will be discussed further in sections III and IV assuming different
distributions for $m(x)$. The reader has perhaps noticed that one could
develop various other expressions for the Landauer conductance that
involve evolving, instead of a two fermion state, a two boson
state or a one boson, one fermion state; but eq (44) is the form\
we shall use here.

Finally let us briefly examine some symmetries of $H_{SUSY}$.
Evidently, it conserves the R fermion charge; that is,
\begin{equation}
[ Q_{F}^{R}, H_{SUSY} ] = 0
\end{equation}
where $ Q_{F}^{R} \equiv c_{+}^{R \dagger} c_{+}^{R} -
c_{-}^{R \dagger} c_{-}^{R} $. Similarly it also conserves
the R boson charge as well as the corresponding quantity
for the A boson and fermion. $ H_{SUSY} $ also possesses
a supersymmetry analogous to that displayed by the 
models studied in refs \cite{zirnbauer,leon1,ilya1,ilya2,kondev,leon2}. 
Specifically the Hamiltonian, eq (42), commutes
with the supercharges $ Q_{FB}^{R}, Q_{FB}^{A}, Q_{BF}^{R} $
and $Q_{BF}^{A}$ which are given by
\begin{eqnarray}
Q_{FB}^{R} & \equiv & c_{+}^{R} b_{+}^{R \dagger} + c_{-}^{R \dagger}
b_{-}^{R}, \nonumber \\
Q_{BF}^{R} & \equiv & c_{+}^{R \dagger} b_{+}^{R} - c_{-}^{R}
b_{-}^{R \dagger};
\end{eqnarray}
$ Q_{FB}^{A} $ and $ Q_{BF}^{A} $ are obtained by replacing
$ R \rightarrow A $ in eq (46).

It will be seen below that after disorder averaging we must analyse 
an effective interacting Hamiltonian instead of the non-interacting
(but random) Hamiltonian of eq (42). The effective interacting 
Hamiltonian will also possess the symmetries discussed above.

\section{Two Edge Chalker-Coddington Model}

\subsection{Disorder Average}

First let us assume that the real and imaginary parts of $ m(x) $
are independent gaussian white noise processes with zero mean.
Thus $ [ m(x) ]_{{\rm ens}} = 0 $. Here $ [ \ldots ]_{{\rm ens}}$
denotes an average over the ensemble of $m(x)$ realizations. 
The variance is given by
\begin{eqnarray}
[ m^{*}(x), m(x') ]_{{\rm ens}} & = & D \delta ( x - x' )
\nonumber \\
{[ m(x), m(x') ]}_{{\rm ens}} & = & 0.
\end{eqnarray}
Essentially the same statistics result if it is assumed
that $ m(x) $ is a rapidly fluctuating phase factor. These
assumptions are similar to those made in the Chalker-Coddington
model. 
For simplicity we suppose $ E = 0$. It is shown in Appendix A
that this entails no loss of generality or modification of behaviour;
but the averaging is simpler to describe. 

To calculate the average Landauer conductance, we must average
the S-matrix $S_{SUSY}(x_{f}, x_{i})$ over the ensemble described
above. This is accomplished by expanding the S-matrix in a formal
series (eq 28) and averaging term by term. The result is
\begin{equation}
[ S_{SUSY}(x_{f}, x_{i})]_{{\rm ens}} = \exp[ - (D/2) H_{CC} (x_{f} - x_{i}) ]
\end{equation}
where
\begin{equation}
H_{CC} \equiv \hat{A} \hat{B} + \hat{B} \hat{A}.
\end{equation}
$\hat{A}$ and $\hat{B}$ are defined below eq (42). The ensemble averaged
conductance is then given by
\begin{equation}
[ g ]_{{\rm ens}} = \frac{e^2}{h} <0|c_{+}^{A} c_{+}^{R} \exp \left( - \frac{D}{2} 
H_{CC} ( x_{f} - x_{i} ) \right) c_{+}^{R \dagger} c_{+}^{A \dagger} |0>.
\end{equation}
Thus in order to calculate the average conductance we need to study the evolution
of a two fermion state under the effective Hamiltonian $H_{CC}$. In contrast to
$H_{SUSY}$, $H_{CC}$ is not random; but the price paid is that it is interacting.

\subsection{Anderson Localization}

The evaluation of eq (50) is simplified by the observation that all of the fermionic
bilinears that appear in $H_{CC}$ (namely $ c_{+}^{R \dagger} c_{-}^{R \dagger},
c_{-}^{R} c_{+}^{R}, c_{+}^{A \dagger} c_{-}^{A \dagger}$ and
$ c_{-}^{A} c_{+}^{A} $) annhilate the two fermion state. Hence the expression
for the average conductance, eq (50), simplifies to 
\begin{equation}
[ g ]_{{\rm ens}} = \frac{ e^2 }{ h } <0|_{B} \exp \left( - \frac{D}{2}
H_{CC}^{{\rm boson}} (x_{f} - x_{i}) \right) |0>_{B}.
\end{equation}
Here $ H_{CC}^{{\rm boson}}$ is the purely bosonic part of $H_{CC}$ (written
explicitly below) and $|0>_{B}$ is the boson vacuum. Physically then we
need only calculate the boson vacuum amplitude.

Now let us analyse the boson Hamiltonian
\begin{eqnarray}
H_{CC}^{{\rm boson}} & = & ( b_{+}^{R \dagger} b_{-}^{R \dagger} - b_{-}^{A} b_{+}^{A} )
( b_{-}^{R} b_{+}^{R} - b_{+}^{A \dagger} b_{-}^{A \dagger} ) \nonumber \\
& & +
( b_{-}^{R} b_{+}^{R} - b_{+}^{A \dagger} b_{-}^{A \dagger} )
( b_{+}^{R \dagger} b_{-}^{R \dagger} - b_{-}^{A} b_{+}^{A} ).
\end{eqnarray}
In contrast to the full Hamiltonian $ H_{CC} $, the bosonic part is 
Hermitean. Note that it is also positive definite since it is of the
form $ \hat{D}^{\dagger} \hat{D} + \hat{D} \hat{D}^{\dagger} $ where
$ \hat{D} \equiv ( b_{-}^{R} b_{+}^{R} - b_{+}^{A \dagger} b_{-}^{A \dagger} ) $.
After some manipulation we obtain the more revealing form
\begin{eqnarray}
H_{CC}^{{\rm boson}} & = & h_{n} + h_{+} + h_{-}; \hspace{2mm} {\rm with}
\nonumber \\
h_{n} & \equiv & 2 n_{+}^{R} n_{-}^{R} + 2 n_{+}^{A} n_{-}^{A} + n_{+}^{R}
+ n_{-}^{R} + n_{+}^{A} + n_{-}^{A} + 2 ; \nonumber \\
h_{+} & \equiv & - 2 b_{+}^{R \dagger} b_{-}^{R \dagger}
b_{+}^{A \dagger} b_{-}^{A \dagger} ;\nonumber \\
h_{-} & \equiv & - 2 b_{+}^{R} b_{-}^{R} b_{+}^{A} b_{-}^{A}.
\end{eqnarray}
Here $ n_{+}^{R} = b_{+}^{R \dagger} b_{+}^{R} $ etc.
Although lengthy, eq (53) has a simple content. Consider the 
bosonic state $ |n>_{B} $, $n = 0,1,2, \ldots $ $|n>_{B}$ is defined
as a normalized state that contains $n$ bosons of each kind (R+, R-, A+
and A-). $n=0$ corresponds to the boson vacuum. Inspection of eq (53)
shows that these states are closed under the action of $ H_{CC}^{{\rm boson}}$.
Hence we need only consider the block of $H_{CC}^{{\rm boson}} $ that
lies within the invariant subspace spanned by these states.

Our plan therefore is to find the eigenstates of $H_{CC}^{{\rm boson}}$ 
that lie within the subspace spanned by $ |n>_{B}$. Expansion of the
boson vacuum in terms of these eigenstates will then allow straightforward
evaluation of the boson vacuum amplitude and the average Landauer 
conductance, eq (51). Note that the Hermiticity of $H_{CC}^{{\rm boson}}$
ensures that it's eigenstates form a complete set and it is therefore
appropriate to use them as a basis.

The coupled boson Hamiltonian, $H_{CC}^{{\rm boson}}$, is solved
in section C below. It is found to have a continuum of eigenstates
denoted $|k>$ with eigenvalue $(1+ k^2)/2$. Here $k \in [0, \infty]$.
The eigenstates are orthogonal and are normalized so that $<k|0>_{B}
=1$ and 
\begin{equation}
< p | k > = \frac{2}{\pi} \cosh^{2} \left( \frac{ \pi k }{2} \right)
\frac{1}{ k \sinh \left( \frac{ \pi k }{2} \right) } \delta ( k - p ).
\end{equation}
It follows from eq (54) and 
the presumed completeness of the eigenfunctions of 
$H_{CC}^{{\rm boson}}$
\begin{equation}
\frac{\pi}{2} \int_{0}^{\infty} d k \frac{ k \sinh \frac{\pi k}{2} }{
\cosh^{2} \frac{ \pi k }{ 2 } } |k><k| = \cal{I}.
\end{equation}
Here $\cal{I}$ denotes the identity matrix in the subspace 
spanned by $ |n>_{B}$.

Inserting the resolution of the identity eq (55) into the conductance
formula, eq (51), yields
\begin{eqnarray}
[ g ]_{{\rm ens}} & = & 
\frac{ \pi e^2 }{ 2 h } \exp \left( - \frac{ D}{4} ( x_{f}
- x_{i} ) \right) 
\nonumber \\
& & \times
\int_{0}^{\infty} d k \frac{ k \sinh \frac{\pi k}{2} }{
\cosh^{2} \frac{ \pi k }{ 2 } } \exp \left( - \frac{ D k^{2} }{4} ( x_{f}
- x_{i}) \right).
\end{eqnarray}
In the limit of large sample size we find
\begin{equation}
[ g ]_{ens} \approx \frac{ e^{2} }{ h } \left( \frac{ \pi }{ D( x_{f} - x_{i} ) }
\right)^{\frac{3}{2}} \exp \left( - \frac{D}{4} ( x_{f} - x_{i} ) \right).
\end{equation}
Thus the model exhibits Anderson localization: the conductance decays
exponentially for sufficiently large sample size as generally expected
of a dirty one dimensional quantum wire. Eq (57) agrees with the result
obtained by ref \cite{zirnbauer}.

\subsection{Solution of Coupled Boson Hamiltonian}

We wish to solve the Schr\"{o}dinger equation 
\begin{equation}
H_{CC}^{{\rm boson}} | \psi > = \lambda | \psi >
\end{equation}
within the subspace spanned by $ |n>_{B} $. Expand the eigenstate
as
\begin{equation}
| \psi > = a_{0} |0>_{B} + a_{1} |1>_{B} + \ldots
+ a_{n} |n>_{B} + \ldots
\end{equation}
The effect of $ H_{CC}^{{\rm boson}} $, eq (53), on the states
$|n>_{B}$ is easily computed:
\begin{eqnarray}
h_{n} |n>_{B} & = & ( 4 n^2 + 4 n + 2 ) |n>_{B};
\nonumber \\
h_{+} |n>_{B} & = & - 2 (n+1)^{2} | n+1 >_{B};
\nonumber \\
h_{+} |n>_{B} & = & - 2 n^{2} | n - 1 >_{B}.
\end{eqnarray}

Eq (59) and (60) together allow us to write the Schr\"{o}dinger
equation as a three term recurrence relation
\begin{equation}
( 4 n^2 + 4 n + 2 ) a_{n} - 2 n^2 a_{n-1} -
2 (n+1)^{2} a_{n+1} = \lambda a_{n}
\end{equation}
subject to $ a_{-1} = 0 $. Our goal now is to
solve eq (61) for different $ \lambda $ and then
ortho-normalize the solutions. The process of
ortho-normalization will weed out the disallowed
values of $ \lambda $. Note that from the Hermiticity
and positive definiteness of $ H_{CC}^{{\rm boson}} $
mentioned following eq (52), we are already assured
that the allowed $ \lambda $ must be positive and
real.

To solve the recurrence relation we introduce the
generating function
\begin{equation}
f(x) = a_{0} + a_{1} x + \ldots + a_{n} x^{n} + \ldots
\end{equation}
From the recurrence relation, eq (61), we can easily
construct the differential equation obeyed by $f$:
\begin{equation}
x (x-1)^{2} \frac{ d^2 }{ d x^2 } f 
+ ( 3 x - 1 )( x - 1 ) \frac{ d }{ d x } f 
+ \left( x - 1 + \frac{ \lambda}{2} \right) f = 0.
\end{equation}
This equation has three regular singular points: at $ x=0, 1 $
and $ \infty $; it is therefore a Riemann P-equation. One solution
is analytic at $ x= 0 $ as can be verified by directly substituting
the series, eq (62), in the differential eq (63). This is the
solution we seek; it generates the solution to eq (61). 

The solution to a Riemann P-equation can always be expressed
in terms of Hypergeometric functions. Making the standard
transformations (see, e.g., ref \cite{morse}, chapter 5) 
we find the analytic solution is given by 
\begin{eqnarray}
f(x) & = & ( 1 - x )^{\mu} F ( \mu + 1, \mu + 1, 1; x );
\nonumber \\
{\rm where} \hspace{3mm} \mu & = & - \frac{ 1 }{2} + \frac{ \sqrt{ 1 - 
2 \lambda } }{2}.
\end{eqnarray}
Note that for $ \lambda > 1/2 $, $ \mu $ becomes complex. Eq (64) 
therefore points to some change in behaviour at $ \lambda = 1/2 $.
Below we will find that in fact the allowed eigenvalues are
$ \lambda > 1/2 $.

The coefficients $ a_{n} $ can be extracted from $ f(x) $ via
the contour integral
\begin{equation}
a_{n} = \oint_{C} \frac{ d x }{ 2 \pi i } \frac{1}{x^{n+1}} f(x).
\end{equation}
Here $C$ is a contour that encircles the origin but not 
the branch point $ x=1 $. This
contour integral cannot be expressed in elementary form in general
but the large $ n $ behaviour of $ a_{n} $ can be obtained
by using an integral representation for the hypergeometric function
(see Appendix B). 

Let us focus on the solutions to eq (61) with $ \lambda > 1/2 $.
Write $ \lambda = 1/2 + k^2/2 $ where $ k \in [ 0, \infty ] $.
The large $n$ asymptotic behaviour is
\begin{equation}
a_{n} \approx \frac{ 2 }{ \pi } \cosh \frac{ \pi k }{ 2 } \frac{1}{ 
\left( 2 k \sinh \frac{\pi k}{2} \right)^{1/2} } \frac{ 1 }{ \sqrt{n} }
\cos \left( \frac{k}{2} \ln n + \phi(k) \right).
\end{equation}
Here the phase $ \phi(k) = \arg [ \Gamma ( i k ) \Gamma( 1/2 - ik/2 ) / \Gamma^{2}
( 1/2 + i k/2 ) ]$. Thus the solutions for $ \lambda > 1/2 $ decay slowly
( as $ 1/\sqrt{n} $ ) and exhibit weak logarithmic oscillations. 

Fortunately, it turns out that this asymptotic behaviour is sufficient to
orthonormalize the eigenstates. Let $ b_{n} $ be a solution to eq (61)
with eigenvalue $ \rho $:
\begin{equation}
( 4 n^2 + 4 n + 2 ) b_{n} - 2 n^2 b_{n-1} -
2 (n+1)^{2} b_{n+1} = \rho b_{n}.
\end{equation}
Multiply eq (61) by $ b_{n} $, eq (67) by $a_{n}$, take the difference
and sum on $ n $. A series of cancellations allows us to express the
sum as a surface term\footnote{This result is not surprising when one
considers the analogous result of Strum-Liouville theory where it is
known that the overlap integral of distinct eigenfunctions can be 
expressed as a surface term. See, for example, ref \cite{morse}, p 719-720.}
\begin{equation}
(\rho - \lambda) \sum_{n=0}^{N} a_{n} b_{n} = 2 (N+1)^{2} (a_{N+1} b_{N} 
- a_{N} b_{N+1} ).
\end{equation}
Thus the asymptotic behaviour of $ a_{n} $ is sufficient to evaluate 
the orthonormalization sum $ \lim_{N \rightarrow \infty} \sum_{n=0}^{N}
a_{n} b_{n} $. Use of eq (66), (68) and the delta function 
representation\footnote{Used for example to obtain the orthonormalization
of plane waves.}
\begin{equation}
\lim_{L \rightarrow \infty} \frac{1}{k} \sin k L = \pi \delta (k)
\end{equation}
yields eq (54).

We have so far focussed on solutions with $ \lambda > 1/2 $. Using 
similar arguments one can show that solutions with $ 0 < \lambda < 1/2$
do not decay fast enough for large $ n $ to be ortho-normalizable 
even as continuum
eigenfunctions. They are analogous to the exponential negative energy
solutions of the free particle Schr\"{o}dinger equation in elementary
quantum mechanics. In the same
spirit, the solutions for $ \lambda > 1/2 $ are analogous to the
positive energy continuum of plane wave solutions.

Having shown that there are no eigenstates with $ \lambda < 1/2 $, we
have justified the step from eq (54) to eq (55). Alternatively
one might try to prove the completeness relation, eq (55), directly.
Forming the matrix element of eq (55) between the states $ |n>_{B} $
and $ |m>_{B} $, one should check whether 
\begin{equation}
\frac{\pi}{2} \int_{0}^{\infty} d k \frac{ k \sinh \frac{ \pi k }{ 2 } }{
\cosh^{2} \frac{ \pi k }{ 2 } } a_{m} (k) a_{n} (k) = \delta_{nm}.
\end{equation}
We have verified this analytically (and more extensively numerically)
for a few small values of $ n $ and $ m $ for which the $a'$s can 
be computed directly from eq (61). It may be possible to
construct the general proof using eq (64-65), but this passes
beyond conventional standards of rigour and good taste in theoretical
physics.

\section{Delocalization Transition in Dyson Glass}

Now let us analyse the case in which $m(x)$ is purely
imaginary. We rewrite eq (2) as
\begin{equation}
- i \frac{ \partial }{ \partial x } 
\left( \begin{array}{c}
\psi_{+} \\
\psi_{-} 
\end{array} \right)
= \left( \begin{array}{cc}
E & - i m \\
- i m & - E 
\end{array}
\right) \left( \begin{array}{c}
\psi_{+} \\
\psi_{-} 
\end{array} \right).
\end{equation}
Here $ m(x) $ is a real white noise process 
with mean $ m_{0} $. We have recycled the symbol
$m$ to conserve the finite resources of the alphabet.
As noted in the introduction,
this model is related to a model of glass first analysed
by Dyson \cite{dyson} and to a special version of the random bond 
Ising Model introduced by McCoy and Wu \cite{mccoy}. It has been 
extensively studied previously in its various incarnations \cite{daniel,leon2}
and it is known to have a critical point at $ m_{0} = 0$
and $ E=0$. Here we shall use the
Landauer conductance to investigate this critical point.
For simplicity, the discussion is limited to two circumstances.
First we set $ E=0 $ and study the behaviour of the conductance
as $ m_{0} \rightarrow 0$. Next we set $m_{0}=0$ and tune the
Fermi energy, $E$.

\subsection{Elementary Solution for $E=0$}

The techniques developed in this paper can be used to 
calculate the disorder averaged Landauer conductance of
this model; but for the special case $E=0$ a much more
complete solution can be obtained by elementary 
means. The content of this solution is instructive
and it is therefore described. 

When $E=0$ the transfer matrix of eq (71) has a particularly
simple explicit form for arbitrary $ m (x) $. Define 
\begin{equation}
M \equiv \int_{x_{i}}^{x_{f}} d x m(x).
\end{equation}
Then the transfer matrix
\begin{equation}
T = \left( \begin{array}{cc}
\cosh M & \sinh M \\
\sinh M & \cosh M 
\end{array} \right).
\end{equation}
Use of eq (4), (5) and (73) immediately yields
\begin{equation}
t = {\rm sech} \hspace{2mm} M
\end{equation}
and
\begin{equation}
g = \frac{e^2}{h} {\rm sech}^2 \hspace{2mm} M.
\end{equation}
Since $m$ is a Gaussian white noise process $M$ is
a Gaussian random variable. If we suppose 
\begin{eqnarray}
[ m ]_{{\rm ens}} & = & m_{0} \hspace{3mm} {\rm and} \nonumber \\
{[ \delta m (x) \delta m (x') ]_{{\rm ens}}} & = & D \delta (x-x'),
\end{eqnarray}
where $ \delta m (x ) \equiv m (x) - m_{0} $, it is easy to
calculate
\begin{eqnarray}
[ M ]_{{\rm ens}} & = & m_{0} ( x_{f} - x_{i} ) \hspace{3mm} {\rm and} 
\nonumber \\
{[ ( \delta M )^2 ]_{{\rm ens}}} & = & D (x_{f} - x_{i}).
\end{eqnarray}
From eq (77) the full distribution of $M$ can be reconstructed:
\begin{equation}
P(M) = \frac{ 1 }{ \sqrt{ 2 \pi D (x_{f} - x_{i}) } } \exp \left(
- \frac{ [ M - m_{0} (x_{f} - x_{i}) ]^{2} }{ 2 D (x_{f} - x_{i}) }
\right).    
\end{equation}
It is now altogether straightforward to obtain statistical information
about the Landauer conductance from eq (75) and (78). 
In particular it can be shown: (i) At 
the critical point the average conductance decays as a power of the 
system size $ (x_{f} - x_{i})^{-1/2} $. (ii) Away from the critical point
$(m_{0} \neq 0)$ the average conductance decays exponentially (Anderson
localization). The localization length diverges as $ m_{0}^{- \nu } $
where the exponent $ \nu = 2 $. (iii) The conductance has a very broad
distribution; the average is dominated by large rare fluctuations. (iv)
Away from the critical point the typical conductance also decays 
exponentially but with a distinct localization length that diverges
with exponent $ \nu = 1 $ as the critical point is approached.

In summary although eq (71) is essentially trivial to solve for $E=0$,
the content of the solution is not trivial. It illustrates many general
features of random critical systems \cite{daniel}.

\subsection{Tuning Fermi Energy}

\subsubsection{Disorder Averaging} 

The localization length is expected to diverge as $ ( \ln E )^{2} $
as $ E \rightarrow 0 $. To probe this singularity we shall make
one simplification. The frequency $E$ will be considered to be complex,
and we shall approach the singularity at the origin along the 
imaginary axis. Thus we shall replace $ E \rightarrow i \omega $
in eq (71) and study $ [ g ]_{{\rm ens}} ( \omega ) $.

Note that for imaginary frequency eq (71) is real and therefore
the scattering amplitude $t$ is also real. In general $t$ is
not a suitable probe of delocalized behaviour as it decays
exponentially (due to phase fluctuations) even when $|t|^2$ does
not. In this case however we may look for the transition by calculating
the average of $t$ which is not only real, but also positive (and
therefore free of sign fluctuations\footnote{ Proof: Consider a series 
of truncated problems as in section II C. First note that for $x_{0}
= x_{f} $, $ t(x_{0}) = 1$. Then observe that $ t ( x_{0} ) $ can
never vanish; if it did, the wave function vanishes in the drain,
and by integrating eq (71) backwards, it vanishes in the sample
and source also. Since it cannot pass through zero, by continuity,
$ t(x_{0}) $ must remain positive for all $ x_{0} $ including 
$x_{i}$. }). This constitutes a simplification because to calculate
$t$ the A fermions and bosons are not needed. 

Even if one insists on calculating $ |t|^{2} $, the A fermions and
bosons are unneeded. Since $ t $ is real $ |t|^2 = t^2 $ and
we may write
\begin{equation}
|t|^2 = <0| c_{+}^{R} S_{F}^{R}( x_{f}, x_{i} ) c_{+}^{R \dagger} 
|0> 
<0| b_{+}^{R} S_{B}^{R}( x_{f}, x_{i} ) b_{+}^{R \dagger} |0>.
\end{equation}

Having noted these possibilities, we now disregard them.
Instead we use eq (44) to calculate $|t|^2 $. Hence we need
to average the S-matrix $S_{SUSY}(x_{f}, x_{i})$ over the white
noise ensemble of real $m(x)$ with $[m(x)]_{{\rm ens}}=0$ and
$[ m(x), m(x') ]_{{\rm ens}} = D \delta (x - x') $. The result
is 
\begin{equation}
[ S_{SUSY} (x_{f}, x_{i}) ]_{{\rm ens}} = \exp \left( -
\frac{D}{2} H_{D} ( x_{f} - x_{i} ) \right)
\end{equation}
where
\begin{equation}
H_{D} \equiv \omega \hat{N} + \frac{D}{2} \hat{K}^{2}.
\end{equation}
$\hat{N}$ counts the total number of particles regardless
of species and $K \equiv i ( c_{+}^{R \dagger} c_{-}^{R \dagger}
+ c_{-}^{R} c_{+}^{R} + c_{+}^{A \dagger} c_{-}^{A \dagger}
+ c_{-}^{A} c_{+}^{A} - b_{+}^{R \dagger} b_{-}^{R \dagger}
+ b_{-}^{R} b_{+}^{R} - b_{+}^{A \dagger} b_{-}^{A \dagger}
+ b_{-}^{A} b_{+}^{A} )$.
The ensemble averaged conductance is then given by
\begin{equation}
[g]_{{\rm ens}} = \frac{e^2}{h} <0| c_{+}^{A} c_{+}^{R} 
\exp \left( - \frac{D}{2} H_{D} ( x_{f} - x_{i} ) \right)
c_{+}^{R \dagger} c_{+}^{A \dagger} |0>.
\end{equation}

\subsubsection{Critical Behaviour}

Once again all the fermionic bilinears in $\hat{K}^{2}$ annhilate
$ c_{+}^{R \dagger} c_{+}^{A \dagger} |0> $; hence the conductance
is determined by the boson vacuum amplitude
\begin{eqnarray}
[ g ]_{{\rm ens}} & =  & \frac{ e^2 }{ h } 
\exp [ - 2 \omega (x_{f} - x_{i}) ] \nonumber \\
 & & 
\times <0|_{B} \exp \left( 
- \frac{D}{2} H_{D}^{{\rm boson}} (x_{f} - x_{i}) \right) |0>_{B}.
\end{eqnarray}
As before $ H_{D}^{{\rm boson}} $ is hermitean and positive definite. 
It has an invariant subspace spanned by the states $ |n>_{B} $, 
$n =0,1,2, \ldots $. $|n>_{B}$ is here defined as the state with
2 $n$ bosons of each kind. Hence we may focus on the block of 
$H_{D}^{{\rm boson}}$ that lies within the subspace spanned by
$|n>_{B}$.

Our plan is to find eigenstates of this block and to expand the
boson vacuum in terms of these eigenstates. In contrast to the
two edge Chalker-Coddington case the eigenstates are discrete
and can be labelled by an integer $l = 1, 2, 3, \ldots $. The eigenvalue problem is solved
approximately in the relevant 
small $ \omega $ limit in the next subsubsection.
It will be found that the eigenvalues are given by
\begin{equation}
\lambda_{l} = \frac{ 2 \pi^2 l^2 }{ ( \ln \frac{\omega}{D} )^2 }
\end{equation}
and the overlap with the vacuum is 
\begin{equation}
|<l|0>_{B}|^{2} = \frac{ 1 }{ | 2 \ln \frac{\omega}{D} | }
\frac{ \sinh \pi k_{l} }{\pi k_{l}}
\end{equation}
where $ k_{l} \equiv l \pi/ \ln( \omega/D ) $.
Putting together these results
\begin{equation}
[ g ]_{{\rm ens}} \approx \frac{ e^2 }{ h } \frac{1}{| 2 \ln \frac{\omega}{D} | }
\exp \left( - 2 \pi^2 D \frac{1}{ (\ln \frac{ \omega }{D} )^2 } (x_{f} - x_{i})
\right).
\end{equation}
for large samples.
Eq (86) shows that the sample is Anderson localized away
from the critical point. The localization length diverges
as $ ( \ln \omega )^2 $ as $ \omega \rightarrow 0 $, in
agreement with previous work (Balents and Fisher \cite{leon2} and refs
therein). 

\subsubsection{Solution of Boson Problem}

This calculation is very similar to the corresponding calculation
by Balents and Fisher \cite{leon2} of the average Green's function . For this 
reason, after pointing out the special feature of the conductance 
calculation (discussion leading to eq (89)) we present the remaining
steps in outline. The reader interested in more details should 
consult the appropriate sections of Balents and Fisher.

We wish to solve 
\begin{equation}
H_{D}^{{\rm boson}} | \psi > = \lambda | \psi > 
\end{equation}
within the subspace spanned by the states $ |n>_{B} $ defined
in the previous subsubsection. To this end expand $|\psi>$ as
\begin{equation}
| \psi > = a_{0} |0>_{B} + a_{1} |1>_{B} + \ldots +
a_{n} |n>_{B} + \ldots 
\end{equation}
A difficulty arises if we assume the states are normalized
so that $ <n|_{B} |m>_{B} = \delta_{mn} $. In this case the
effect of $H_{D}^{{\rm boson}} $ on $|n>_{B}$ is to yield a
linear combination of $ |n+1>_{B}$, $|n>_{B}$ and $|n-1>_{B}$
with coefficients that involve products such as $ \sqrt{ n (n+1)}$.
Such non-integer 
coefficients would then also appear when eq (87) is written
as a recurrence relation analogous to eq (61) and would defeat
the generating function technique that was used to solve eq (61).

However if we work with an unusual normalization such that
$ <n|_{B} |m>_{B} = (n+1) \delta_{mn} $ simple coefficients result.
Adopting this convention we find
\begin{eqnarray}
\left( \frac{4 \omega}{D} m + 4 m^2 + 4 m + 1 \right) a_{m} & &
\nonumber \\
- m(2m-1) a_{m-1} - (m+1)(2m+3) a_{m+1} = \lambda a_{m}. & &
\end{eqnarray}  
The corresponding generating function obeys a differential
equation with three singular points, two regular and one 
irregular. Thus the solutions are related not to the common
functions of mathematical physics but to the more obscure
Mathieu or Spheroidal functions. Rather than pursue this 
direction we closely follow Balents and Fisher to obtain
an approximate solution to the Schr\"{o}dinger eq (89), 
valid in the interesting limit of small $ \omega/D $.

First we solve eq (89) for $ \omega = 0 $ using the generating
function method. This solution should be accurate for $ n \ll
D/\omega $. Setting $ a_{0} = 1 $ we find
\begin{equation}
a_{n} \approx \frac{1}{2n} \left( \frac{ \sinh \pi k }{ \pi k }
\right)^{1/2} \cos [ k \ln n - \phi( k ) ]
\end{equation}
for $ 1 \ll n \ll D/\omega $. Here we have introduced $ k $ via
$ \lambda = 2 k^2 $ and $ \phi(k) \equiv \arg \Gamma( 1 - ik) +
2 k \ln 2 $. Next we approximate the large $n$
behaviour by taking the continuum limit of eq (89):
\begin{equation}
n^2 \frac{ d^2}{dn^2} a + 3 n \frac{ d }{ dn} a 
+ \left( \frac{ \lambda }{2} + 1 - \frac{ \omega }{D} n \right) a = 0.
\end{equation}
Eq (91) is soluble by Laplace transforms; but an even quicker solution
may be effected by introducing $ f( n ) \equiv n a( n ) $ which obeys
the equation
\begin{equation}
n^2 \frac{ d^2 }{ d n^2 } f + n \frac{ d }{ d n } f 
+ \left( \frac{ \lambda }{ 2 } - \frac{ \omega }{ D } n \right) f = 0.
\end{equation}
Precisely this equation was analysed by Balents and Fisher in Appendix
D of their paper. Borrowing their results we can find the values of
$ \lambda $ for which eq (91) has a solution that decays as $ n \rightarrow 
\infty $ and smoothly matches eq (90) for $ 1 \ll n \ll D/\omega $. This
yields the quantized energy levels of eq (84). 

The eigenstates may be approximately normalized by taking the continuum
solution to apply everywhere. The normalization sum $ \sum_{ n = 0 }^{\infty}
a_{n}^{2}/(n+1) $ may then be replaced by an integral. The relevant integral
is evaluated asymptotically by Balents and Fisher (Appendix D). Transcription
of their result yields eq (85).

\subsection{Summary}

For reference we summarize the results of this section:

At the critical point $(m_{0}=0, \omega = 0$) the average conductance
decays as the inverse square root of the sample size.  Off the
critical point the average conductance decays exponentially on a length
scale called the localization length. For $ \omega = 0 $ we find the
localization length of the average conductance diverges as $ m_{0}^{-\nu}$
as $ m_{0} \rightarrow 0 $, with exponent $ \nu = 2 $; for $ m_{0} = 0 $,
we find it diverges as $ ( \ln \omega )^2 $. These results for the
average conductance localization length are consistent with previous
calculations of other non-local correlation functions for these
models \cite{daniel,leon2}. 

For $ \omega = 0 $, the model is completely soluble by elementary
methods. In this simple case it shows many of the features generally
expected of random critical systems \cite{daniel}. In particular the
Landauer conductance is very broadly distributed and the average is
dominated by large, rare fluctuations. Away from the critical point
the typical conductance also decays exponentially but with a localization
length different from the average conductance. The localization length for
the typical conductance diverges with exponent 1 as $m_{0} \rightarrow 0$.

\acknowledgments

It is a pleasure to thank Leon Balents, Matthew Fisher, Steve Girvin,
Phil Taylor and Tanmay Vachaspati for encouragement, illuminating 
discussions and telling me about their work prior to publication.
The author is supported by an Alfred P. Sloan Research Fellowship. 

\appendix

\section{}

The purpose of this Appendix is to show that the results of section III
are not modified when $ E \neq 0 $. This can be done in various ways.
The method followed here is chosen because it also illustrates how
to average the S-matrix when the random Hamiltonian, $H_{SUSY}$,
is composed of two non-commuting pieces. 

Factorise the S-matrix as
\begin{equation}
S_{SUSY}(x, x_{i}) = \exp [ i E \hat{M} (x - x_{i}) ] U(x, x_{i}).
\end{equation}
Then eq (42-43) imply that $U$ obeys
\begin{equation}
- i \frac{ \partial}{ \partial x } U(x, x_{i}) =
[ m (x) \hat{A}(x) + m^{*}(x) \hat{B} ] U(x,x_{i})
\end{equation}
and the initial condition $U(x_{i},x_{i}) = 1$. Here
\begin{eqnarray}
\hat{A}(x) & \equiv & \exp[ - i E \hat{M} (x-x_{i}) ] 
\hat{A} \exp[ i E \hat{M} (x-x_{i}) ],
\nonumber \\
\hat{B} (x) & \equiv & \exp[ - i E \hat{M} (x-x_{i}) ] \hat{B}
\exp[ i E \hat{M} (x-x_{i}) ].
\end{eqnarray}
In this interaction representation it is easy to show
\begin{eqnarray}
\hat{A}(x) & = & \exp [ - i 2 E (x-x_{i}) ] \hat{A}, \nonumber \\
\hat{B} (x) & = & \exp[ i 2 E (x-x_{i}) ] \hat{B}.
\end{eqnarray}

The disorder average of $U$ can be computed as in section III.
\begin{equation}
[ U(x, x_{i} ) ]_{{\rm ens}} = \exp \left( - \frac{D}{2} (x-x_{i}) 
( \hat{A} \hat{B} + \hat{B} \hat{A} ) \right).
\end{equation}
Recall $ \hat{A} \hat{B} + \hat{B} \hat{A} \equiv H_{CC} $ (eq 49).
Hence
\begin{eqnarray}
[ S_{SUSY} (x, x_{i}) ]_{{\rm ens}} & =  & \exp[ i E \hat{M} (x - x_{i}) ]
\nonumber \\
& & \times
\exp \left( - \frac{D}{2} H_{CC} ( x - x_{i} ) \right).
\end{eqnarray}

Finally note
\begin{equation}
\hat{M} c_{+}^{R \dagger} c_{+}^{A \dagger} |0> = 0.
\end{equation}

Eq (44), (A6) and (A7) together show that
\begin{equation}
[ g]_{{\rm ens}} = \frac{e^2}{h} <0| c_{+}^{A} c_{+}^{R} 
\exp \left( - \frac{D}{2} H_{CC} (x_{f} - x_{i} ) \right)
c_{+}^{R \dagger} c_{+}^{A \dagger} |0>
\end{equation}
---independent of $E$ as claimed in the paper.

\section{}

The contour integral in eq (65) is evaluated here asymptotically
for large $n$. $ f(x) $ is defined in eq (64). As in the paper,
we focus on $ \lambda > 1/2$ and write $ \lambda = (1+ k^2)/2 $.
Thus $ \mu = -1/2 + i k/2 $. 

Use the integral representation of the hypergeometric function
\begin{equation}
F(a,b,c;x) = \frac{ \Gamma (c) }{ \Gamma(b) \Gamma( c -b ) }
\int_{1}^{\infty} d t ( t - x )^{-a} t^{a-c} (t-1)^{c-b-1}
\end{equation}
valid provided $ {\rm Re} c > {\rm Re b} > 0$ (ref \cite{morse}, chapter 5).

Substitute eq (B1) in eq (65) with $ a \rightarrow \mu + 1 $, $ b 
\rightarrow \mu + 1 $ and $ c \rightarrow 1 $. Exchange the order
of the $ t $ and $ x $ integrals to obtain
\begin{equation}
a_{n} = \frac{ \Gamma(c) }{ \Gamma (b ) \Gamma( c - b ) } \int_{1}^{\infty}
d t t^{a-c} (t-1)^{c-b-1} f_{n}(t)
\end{equation}
where
\begin{equation}
f_{n} ( t ) = \oint_{C} \frac{ d x }{ 2 \pi i } \frac{ 1 }{ x^{n+1} }
(t - x)^{-1/2-ik/2} (1 - x)^{-1/2+ik/2}.
\end{equation}
Here $C$ is a contour that encircles the origin but not the branch
point at $ x=1 $. 

Note that the integrand in eq (B3) has branch points at $ x=1 $
and $ x=t $ where $t$ is some point on the positive real axis
to the right of 1 (and to be eventually integrated over the 
range from 1 to $\infty$). To be consistent with the conventions
of the integral representation, eq (B1), we draw branch cuts 
along the positive real axis from 1 to $+\infty$ and $ t $
to $ + \infty $. Also the phase of $(t-x)$ and $(1-x)$ must
both be taken to be zero when $x$ lies on the real axis to
the left of 1.

The contour $C$ is now deformed to pass above and below the branch
cut. It is closed by a small circle around $x=1$ and by a big
circle at $\infty$. The contribution of the circles to the
contour integral vanishes. The contribution from integrating
above and below the branch cut is
\begin{equation}
f_{n}(t) = \frac{1}{\pi} \cosh \frac{ \pi k }{ 2 }
\int_{1}^{t} d x \frac{1}{x^{n+1}} (t - x)^{-1/2-ik/2}
(x - 1)^{-1/2+ik/2}.
\end{equation}
Once again exchange the order of the $x$ and $t$ integrals
to obtain
\begin{eqnarray}
a_{n} & = & \frac{1}{\pi^2} \cosh^2 \frac{ \pi k }{ 2 }
\int_{1}^{\infty} d x \frac{1}{x^{n+1}} (x-1)^{-1/2+ik/2} 
\nonumber \\
\times & &  
\int_{x}^{\infty} d t t^{-1/2+ik/2} (t-1)^{-1/2-ik/2}  
(t-x)^{-1/2-ik/2}.
\end{eqnarray}
The $ \Gamma $ function prefactors in the integral representation, eq (B1),
have been simplified by use of the formula $ \Gamma(z) \Gamma(1-z)
= \pi/\sin (\pi z) $.

Rescale the $t$ integral so the lower limit becomes $ 1 $. Upon
comparison with the integral representation, eq (B1), it is seen
\begin{eqnarray}
a_{n} & = & \frac{1}{\pi} \cosh \frac{ \pi k }{ 2 } \int_{1}^{\infty}
d x \frac{ 1 }{ x^{n+1}} (x - 1)^{-1/2+ik/2} x^{-1/2-ik/2}
\nonumber \\ 
 & & \times 
F(\frac{1}{2} + i \frac{k}{2}, \frac{1}{2} + i \frac{k}{2},
1; \frac{1}{x} ).
\end{eqnarray}
Since we are interested in large $n$ behaviour introduce $ s = \ln x $
and obtain
\begin{eqnarray}
a_{n} & = & \frac{1}{\pi} \cosh \frac{ \pi k }{ 2 } \int_{0}^{\infty}
d s e^{-ns} ( e^{s} - 1 )^{-1/2 + ik/2} (e^s)^{-1/2-ik/2}
\nonumber \\
& & \times 
F( \frac{1}{2} + i \frac{k}{2}, \frac{1}{2} + i \frac{k}{2}, 1;
e^{-s} ).
\end{eqnarray}

Upto this point all the manipulations have been exact. Eq (B7) shows
that to obtain the large $n$ asymptotic behaviour only the small
$s$ behaviour of the various factors in the integrand is needed.
The hypergeometric function has a simple expansion about zero 
(the hypergeometric series); however we need to expand about one
(since $ s \approx 0 $ implies $ e^{-s} \approx 1 $). This is
accomplished by use of the joining formula (ref \cite{morse})
\begin{eqnarray}
F(a,b,c;x) & = &
\frac{ \Gamma(c) \Gamma(c - a -b)}{ \Gamma( c - a ) \Gamma(c-b)}
F(a,b,a+b-c+1; 1-x) 
\nonumber \\
& &
+ \frac{ \Gamma(c) \Gamma(a+b-c) }{ \Gamma(a) \Gamma(b) } 
( 1 - x)^{c-a-b}
\nonumber \\
& & \times
F(c-a, c-b, c-a-b+1; 1-x ). 
\end{eqnarray}
We obtain
\begin{eqnarray}
F( \frac{1}{2} + i \frac{k}{2}, \frac{1}{2} + i \frac{k}{2}, 1;
e^{-s} ) & \approx & 
\frac{ \Gamma( - i k ) }{ \Gamma^{2} ( 1/2 - ik/2 ) }
\nonumber \\
& &
+ \frac{ \Gamma (ik) }{ \Gamma^{2} ( 1/2 + i k/2 ) } s^{-ik}.
\end{eqnarray}
Substitute eq (B9) in eq (B7). This leads to the result of eq (66).

\end{multicols}

\end{document}